**Activation cross section data of deuteron induced nuclear reactions on rubidium up to 50 MeV**

Ferenc Tárkányi[1], Alex Hermanne[2], Ferenc Ditrói[1*], Sándor Takács[1] Anatolij V. Ignatyuk [3], Ingo Spahn [4], Stephan Spellerberg[4]

[1] *Institute for Nuclear Research, (ATOMKI), Debrecen, Hungary*
[2] *Cyclotron Laboratory, Vrije Universiteit Brussel (VUB), Brussels, Belgium*
3 *Institute of Physics and Power Engineering (IPPE), Obninsk, Russia*
4. *Forschungszentrum Jülich, Institute of Neuroscience and Medicine, Nuclear Chemistry (INM-5), Jülich, Germany*

## Abstract

Activation cross sections of the $^{nat}Rb(d,xn)^{87m,85m,85g,83,82}Sr$, $^{nat}Rb(d,x)^{86,84,83,82m}Rb$ and $^{nat}Rb(d,x)^{85m}Kr$ nuclear reactions have been measured for the first time through an activation method combining the stacked foil irradiation technique and gamma-ray spectrometry. The provided cross sections from the present investigation are all new, in such a way contribute to the completeness of the experimental database. The experimental cross sections were compared with the theoretical prediction in the TENDL-2019 TALYS based library and with our calculation using ALICE-D and EMPIRE-D model codes in order the improve their predictivity. Thick target production yields were calculated form the new cross sections for all investigated radioisotopes. Practical applications of the results are shortly discussed.

* Corresponding author: ditroi@atomki.hu

## 1. Introduction

Integral excitation functions for the production of residual nuclides through light charged particle activation constitute basic data for various applications. Nowadays, deuteron induced reactions start to play an important role as the stripping process generates high production yields and the stopping power for deuterons is relatively low. Some years ago, to meet requirements of practical applications, we started to establish an experimental activation database by performing new experiments and a systematic survey of existing data of deuteron induced cross-sections up to 50 MeV. This database is essential for accelerator and target technology to produce high-energy, high-intensity neutron fluxes, wear and material studies through TLA (Thin Layer Activation), medical radioisotope production, space applications (resistance of electronics, shielding, etc.), monitoring of deuteron beam intensities and energies, etc.

Unlike for proton induced reactions, the status (reliability and completeness) of the experimental data for deuteron induced reactions was rather poor, especially above 15-20 MeV. We embarked hence on a systematic program of extending the study to more targets and up to 50 MeV incident deuteron energy. In this work we present the excitation functions for reactions on rubidium (stable isotopes $^{85}$Rb and $^{87}$Rb) leading to radionuclides of Sr, Rb and Kr.

The measured excitation functions are compared with the results of three nuclear reaction model codes. These comparisons can show the present status of the predictivity of these codes and contribute to their development. The reliability of the presently used theoretical codes for deuteron induced reactions is low, compared to proton and alpha particle induced reactions, due to the modeling problems of the deuteron stripping and pickup.

Concerning activation cross sections of deuteron induced reactions on Rb, no experimental data were found in the literature. Only integral physical yields at 22 MeV deuteron energy were reported by Dmitriev et al. [1] for production of $^{84,86}$Rb and $^{85}$Sr.

## 2. Experiments and data evaluation

The main experimental parameters and data evaluation methods used in this work are shown in Table 1. The cross section data were measured by using the activation method, stacked foil irradiation technique and high resolution gamma-ray spectrometry. Beam current and energy scale are based on monitor reactions, re-measured simultaneously over the whole covered energy range (Fig. 1).

The Rb-targets were obtained by deposition of $Rb_2SO_4$ (99.8%, Sigma Aldrich) using a sedimentation method [2] on a 50 μm thick, high purity Al backing in an 11 mm diameter spot. Actual thickness of the layer was determined by measuring the area and differential weighing of the material deposited on the Al backing. The pellets were covered by a 10 μm Al foil for protection. The metal foils were high purity products from Goodfellow.

The stack was irradiated with deuterons of 50 MeV incident energy at an external beam line of the Cyclone 110 cyclotron of the Université Catholique in Louvain la Neuve (LLN): (protons 20-80 MeV; d, α 2.3-27 MeV/nucleon; heavy ions, radioactive ions).

The present stack contained 19 blocks of Al-$Rb_2SO_4$-Al targets alternated with Al-$RuCl_2$-Al targets and further 6 blocks of Al-$Rb_2SO_4$-Al targets alternated with Al-$RuCl_2$-Al targets combined with 20 μm Ti monitors. The nominal current during the 40 min irradiation was 27 nA. The targets were mounted in a target holder (Faraday cup like) provided with a long collimator defining a 5 mm beam diameter.

Gamma-spectra were measured with Canberra GX1520 HPGe detectors of 15% efficiency and 1.9 keV resolution at the 1332 keV $^{60}$Co peak at the VUB (Free University of Brussels) laboratory. The detectors were vertically arranged with adjustable distance sample holders from the detector surface to several meters. The detector head was put into 10 cm lead shielding with 2 mm copper internal lining. The individual foils or a group of target foils were measured in such a distance that the dead time was always under 5%. To make the evaluation easier not more than 5 distances were selected. The samples were measured first after a short cooling time of several hours for the short-lived components up to after several months for the longer-lived radioisotopes. The measuring time was chosen to acquire good statistics for the expected gamma-peaks, but of course it cannot be optimized for every radioisotope in question. The used decay data, taken from the online version of NUDAT 2.6 [3], and the energy threshold values in MeV, obtained from the Q-value calculator [4], are presented in Table 1.

The median beam energies in the individual targets were preliminarily determined by a degradation calculation (see [5]) and were corrected on the basis of the fitted monitor reaction [6]. The activities of the individual radioisotopes were determined from the measurement of the corresponding gamma-peaks (Eq. 1):

$$A_{EOB} = \frac{T}{\varepsilon I_\gamma t_{live}} \frac{\lambda t_{real}}{1 - e^{-\lambda t_{real}}} e^{-\lambda t_c} \qquad \textit{Equation 1}$$

where $A_{EOB}$ is the activity at the End of Bombardment, $T$ is the net area of the gamma-peak, λ is the decay constant, $t_{real}$, $t_{live}$ and $t_c$ are the real-time, live-time of the measurement and the cooling time, $I_\gamma$ is the gamma-line abundance (intensity) and ε is the detector efficiency. From the measured activities the corresponding cross section was calculated according to Eq. 2.

$$\sigma(E) = \frac{A_{EOB} M z e}{I(1 - e^{-\lambda t_i}) \varrho s N_A v f} \qquad \textit{Equation 2}$$

where $z$ is the projectile charge, $I$ *(A)* is beam current, $N_A$ is Avogadro's number ($6.02214\times10^{23}$ mol$^{-1}$), *M (g/mol)* should be molar mass of the chemical compound used as the target material, $e$ is the electron charge ($1.6\text{x}10^{-19}$ C), $\rho$ *(g/cm$^2$)* is the mass density of the target material, *s (cm)* is the thickness of the measured target foil, $v$ is the number of the particular element atoms in the compound molecule and $f$ is the abundance of the particular isotope (entering the particular nuclear reaction) in the element. $\sigma(E)$ *(cm$^2$)* is the cross section at bombarding energy $E$, where $E$ is the median energy in the measured foil of the stack. By using both Eq. 1 and Eq. 2 one must take care of the measuring units. Eq. 2 is valid for isotopic cross sections. In our case mainly elemental cross sections were determined, where *f=1*.

Uncertainties of the median energies were obtained taking into account cumulative effects of possible uncertainties (primary energy, target thickness, energy straggling, correction to monitor reaction). Uncertainty of cross-sections was determined by considering the sum in quadrature of all independent contributions following the ISO recommendations (cf. [7]): beam current (7 %), target thickness (6 %), detector efficiency (5 %), nuclear decay data (3 %), full energy peak area determination and counting statistics (1-20 %).

Natural Rb has two stable isotopes: $^{85}Rb$ (72.17 %) and $^{87}Rb$ (27.83 %). The experimental parameters and the data evaluation methods are summarized in Table 1.

Table 1. Main experimental parameters and data evaluation

| **Experimental parameters** | | **Data evaluation** | |
|---|---|---|---|
| **Incident particle** | deuteron | **Gamma-spectra evaluation** | Genie 2000 [8]<br>Forgamma [9] |
| **Method** | Stacked foil | **Determination of beam intensity** | Faraday cup (preliminary)<br>Fitted monitor reaction (final) [10] |
| **Target and thickness ($mg/cm^2$)** | $Rb_2SO_4$, 99.8 %<br>sedimented 2.7- 45.7 | **Decay data** | NUDAT 2.6 [3] |
| **Number of $Rb_2SO_4$ target samples** | 25 | **Reaction Q-values** | Q-value calculator [4] |
| **Target composition and thickness (μm)** | Al(10)$Rb_2SO_4$(126.5-7.4)Al(50), Al(10)$RuCl_3$ (134.8-16.9)Al(50) repeated 19 times +<br>Ti (22),<br>Al(10)$Rb_2SO_4$(126.5-7.4)Al(50), Al(10) $RuCl_3$ (134.8-16.9)Al(50) repeated 6 times | **Determination of beam energy** | Andersen (preliminary) [5]<br>Fitted monitor reaction (final) [6] |
| **Accelerator** | Cyclone 90 cyclotron of the Université Catholique in Louvain la Neuve (LLN) | **Uncertainty of energy** | Cumulative effects of possible uncertainties |
| **Primary energy (MeV)** | 50 | **Cross sections** | Elemental cross section |
| **Energy range (MeV)** | 49.6-7.9 | **Uncertainty of cross sections** | Sum in quadrature of all individual linear contributions [7] |
| **Irradiation time (min)** | 40 | **Yield** | Physical yield [11, 12] [12] |
| **Beam current (nA)** | 27 | | |
| **Monitor reactions, [recommended values]** | $^{27}Al(d,x)^{22,24}Na$ [6] | | |
| **Monitor target and thickness (μm)** | $^{nat}Al$(50+10) | | |
| **detector** | HPGe | | |
| **γ-spectra measurements** | 4 series | | |
| **Cooling times after EOB (h)** | 5.4-9.0<br>21.8-29.4<br>31.4-125.8<br>240.0-600.8 | | |

Table 2. Decay and nuclear characteristics of the investigated reaction products, contributing reactions and their Q-values

| Nuclide Isomeric level (keV) | Half-life | Decay mode (%) | $E_\gamma$(keV) | $I_\gamma$(%) | Contributing process | Q-value (keV) |
|---|---|---|---|---|---|---|
| $^{87m}$Sr<br>388.533 | 2.815 h | ε: 0.3<br>IT: 99.7 | 388.531 | 82.19 | $^{87}$Rb(d,2n) $^{87m}$Sr | -2724.638 |
| $^{85m}$Sr<br>238.79 | 67.63 min | ε: 13.4<br>IT: 86.4 | 151.194<br>231.860 | 12.8<br>83.9 | $^{85}$Rb(d,2n) $^{85m}$Sr<br>$^{87}$Rb(d,4n) $^{85m}$Sr | -4071.0<br>-22644.0 |
| $^{85}$Sr<br>9/2+ | 64.849 d | ε: 100 | 514.0048 | 96 | $^{85}$Rb(d,2n) $^{85m}$Sr<br>$^{87}$Rb(d,4n) $^{85m}$Sr | -4071.0<br>-22644.0 |
| $^{83m}$Sr<br>259.15 | 4.95 s | IT: 100 | | | $^{85}$Rb(d,4n) $^{83}$Sr<br>$^{87}$Rb(d,6n) $^{83}$Sr | -24519.0<br>-43092.0 |
| $^{83}$Sr | 32.41 h | ε: 100 | 381.53<br>418.37<br>762.65 | 14.0<br>4.2<br>26.7 | $^{85}$Rb(d,4n) $^{83}$Sr<br>$^{87}$Rb(d,6n) $^{83}$Sr | -24519.0<br>-43092.0 |
| $^{82}$Sr | 25.35 d | ε: 100 | 776.511<br>(through<br>$^{82g}$Rb decay) | 15.1 | $^{85}$Rb(d,5n) $^{83}$Sr<br>$^{87}$Rb(d,7n) $^{83}$Sr | -33378.0<br>-51951.0 |
| $^{86m}$Rb<br>556.07 | 1.017 min | IT: 100 | | | $^{85}$Rb(d,p) $^{86}$Rb<br>$^{87}$Rb(d,p2n) $^{86}$Rb | 6426.41<br>-12146.68 |
| $^{86}$Rb | 18.642 d | ε: 0.0052<br>$β^-$: 99.9948 | 1077.0 | 8.64 | $^{85}$Rb(d,p) $^{86}$Rb<br>$^{87}$Rb(d,p2n) $^{86}$Rb | 6426.41<br>-12146.68 |
| $^{84m}$Rb<br>463.59 | 20.26 min | IT: 100 | | | $^{85}$Rb(d,p2n) $^{84}$Rb<br>$^{87}$Rb(d,p4n) $^{84}$Rb | -12704.2<br>-31277.3 |
| $^{84}$Rb | 32.82 d | ε: 96.1<br>$β^-$: 3.9 | 81.6041 | 68.9 | $^{85}$Rb(d,p2n) $^{84}$Rb<br>$^{87}$Rb(d,p4n) $^{84}$Rb | -12704.2<br>-31277.3 |
| $^{83}$Rb | 86.2 d | ε: 100 | 520.3991<br>529.5945<br>552.5512 | 45<br>29.3<br>16.0 | $^{85}$Rb(d,p3n) $^{83}$Rb<br>$^{87}$Rb(d,p5n) $^{83}$Rb<br>$^{83}$Sr decay | -21463.9<br>-40037.0<br>-24519.0 |
| $^{82m}$Rb<br>69.015 | 6.472 h | ε: 100 | 554.35<br>619.11<br>698.37<br>776.52<br>827.83<br>1044.08<br>1317.43<br>1474.88 | 62.4<br>37.98<br>26.3<br>84.39<br>21.0<br>32.07<br>23.7<br>15.5 | $^{85}$Rb(d,p4n) $^{84}$Rb<br>$^{87}$Rb(d,p6n) $^{82}$Rb | -32418.0<br>-50991.0 |
| $^{82}$Rb | 1.2575 min | $β^+$: 95.4<br>EC:4.6 | 776.511 | 15.1 | $^{85}$Rb(d,p4n) $^{81}$Rb<br>$^{87}$Rb(d,p6n) $^{81}$Rb<br>$^{82}$Sr decay | -32418.0<br>-50991.0<br>-33378.0 |
| $^{85m}$Kr<br>304.872 | 4.480 h | IT 21.2<br>$β^-$:78.8 | 304.87<br>151.195 | 14.0<br>75.2 | $^{85}$Rb(d,2p) $^{85m}$Kr<br>$^{87}$Rb(d,2p2n) $^{85m}$Kr | -2129.2<br>-20702.3 |
| $^{85}Kr$ | 10.739 y | β-: 100 | 513.997 | 0.434 | $^{85}Rb(d,2p)\,^{85m}Kr$<br>$^{87}Rb(d,2p2n)\,^{85m}Kr$ | -2129.2<br>-20702.3 |

The Q-values refer to formation of the ground state. In case of formation of a higher laying isomeric state it should be corrected with the energy of level energy of the isomeric state shown in Table 2. When complex particles are emitted instead of individual protons and neutrons the Q-values have to be decreased by the respective binding energies to get the threshold energies (pn→d +2.2 MeV, p2n→t +8.5 MeV, 2pn→$^{3}He$ +7.7 MeV, 2p2n→α +28.3 MeV).

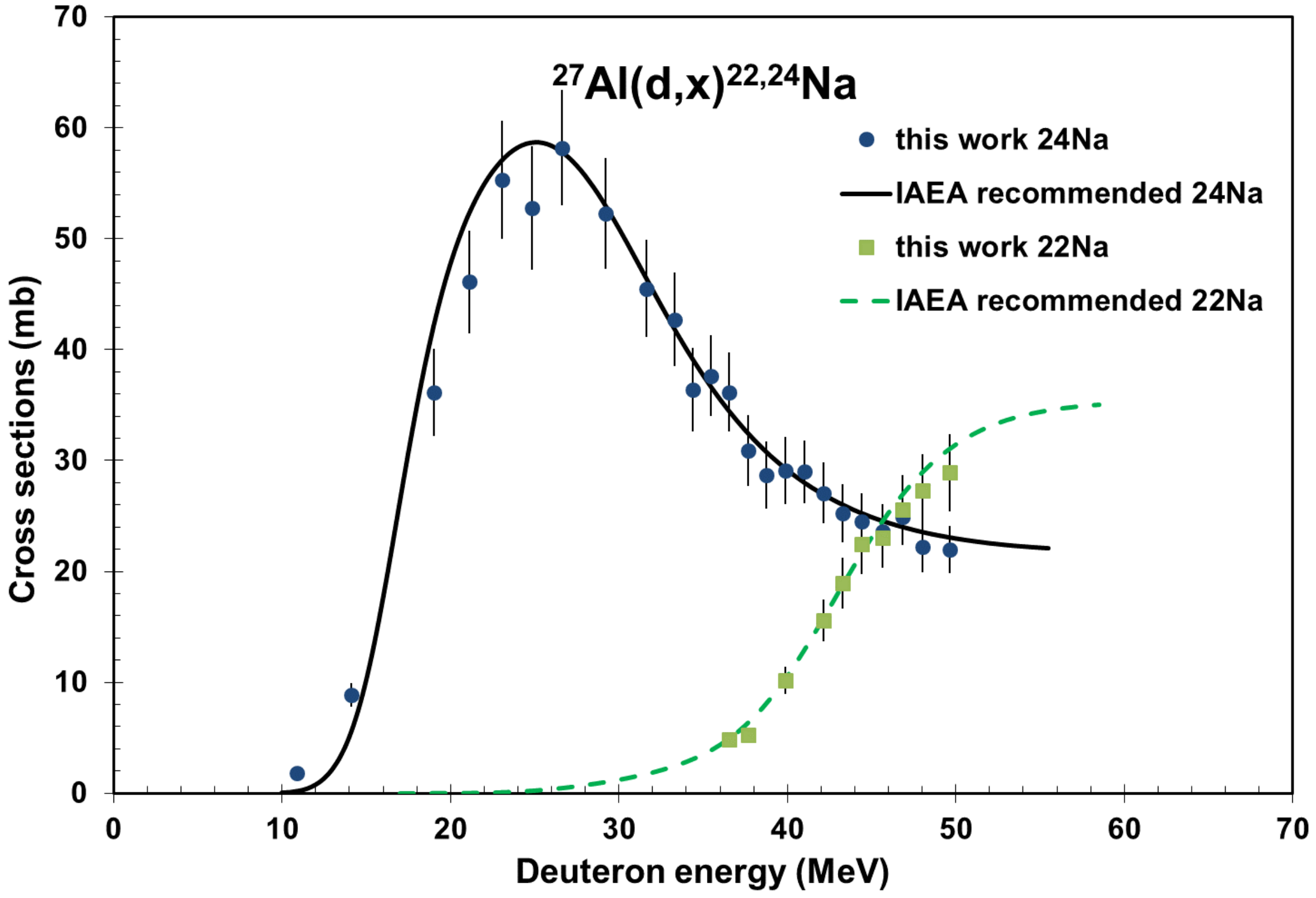

Fig. 1. Excitation function of the simultaneously measured $^{27}Al(d,x)^{24,22}Na$ monitor reactions.

## 3. Nuclear model calculation

The cross sections of the investigated reactions were calculated using the pre-compound model codes ALICE-IPPE [13] and EMPIRE-II [14]. The experimental data are also compared with the cross section data in the TENDL-2019 on-line library [15]. The TENDL-2019 library contains the results of the blind and adjusted calculations on the basis of the latest version (1.6) of the TALYS nuclear model code system [16]. For showing the evolution of this code and database also the results of the TENDL-2017 version are indicated and nearly no or only marginal changes in the results can be noticed.

In our previous works ALICE-IPPE and EMPIRE-II were used successfully for the description of a large amount of reaction cross sections induced by light charged particles. However, during the recent analyses of the (d,p) reactions on the isotopes, $^{114}Cd$ [17]; $^{169}Tm$ [18]; $^{192}Os$ [19] and some others, we were confronted with a large underestimation of the measured cross sections. We come to the conclusion that the experimentally observed cross sections of the (d,p) reaction cannot be reproduced below 20–30 MeV with the available statistical model codes. To achieve a better description of available data for (d,p) reactions with the ALICE and EMPIRE a phenomenological simulation of direct (d,p) and (d,t) transitions were introduced to the above-mentioned codes. A phenomenological enhancement factor K in these relations was taken as energy dependent and estimated to describe the whole set of the observed (d,p) cross sections for medium and heavy nuclei. By this improvement, in the ALICE IPPE-D and EMPIRE-D code versions for the deuteron induced reactions, the direct (d,p) channel is increased strongly and this is reflected in changes for all other reaction channels in both codes. As ALICE-IPPE calculates only the total cross section, for estimation of isomeric states from the ALICE code the isomeric ratios calculated by EMPIRE-D were applied.

## 4. Results and discussion

Formation cross sections for the $^{nat}$Rb(d,xn)$^{87m}$Sr, $^{nat}$Rb(d,xn)$^{85m}$Sr, $^{nat}$Rb(d,xn)$^{85g}$Sr(m+), $^{at}$Rb(d,xn)$^{83g}$Sr(m+), $^{nat}$Rb(d,xn)$^{82}$Sr, $^{nat}$Rb(d,x)$^{86g}$Rb(m+), $^{nat}$Rb(d,x)$^{84g}$Rb(m+), $^{nat}$Rb(d,x)$^{83}$Rb(cum), $^{nat}$Rb(d,x)$^{82m}$Rb and $^{nat}$Rb(d,x)$^{85m}$Kr reactions are presented in Figs. 2–11 in comparison with the TALYS predictions in TENDL-2019 (and TENDL-2017) and with our calculation using the ALICE-D and EMPIRE-D model codes.

### 4.1 Cross sections for production of strontium radioisotopes

The radioisotopes of strontium are produced via direct (d,xn) reactions on one or both of the stable Rb isotopes.

#### 4.1.1 *$^{nat}$Rb(d,xn)$^{87m}$Sr*

The isotope $^{87}$Sr has a stable ground state and a rather short-lived isomeric state $^{87m}$Sr ($T_{1/2}$ = 2.815 h) that was measured with poor statistics because of the long cooling time. It is produced in our experimental circumstances only through the $^{87}$Rb(d,2n) reaction. The agreement of our indicative results (large uncertainties, huge difference in decay corrections between different target foils) with predictions of the theoretical codes is acceptable (Fig. 2). There is no visible difference between the two versions of TENDL.

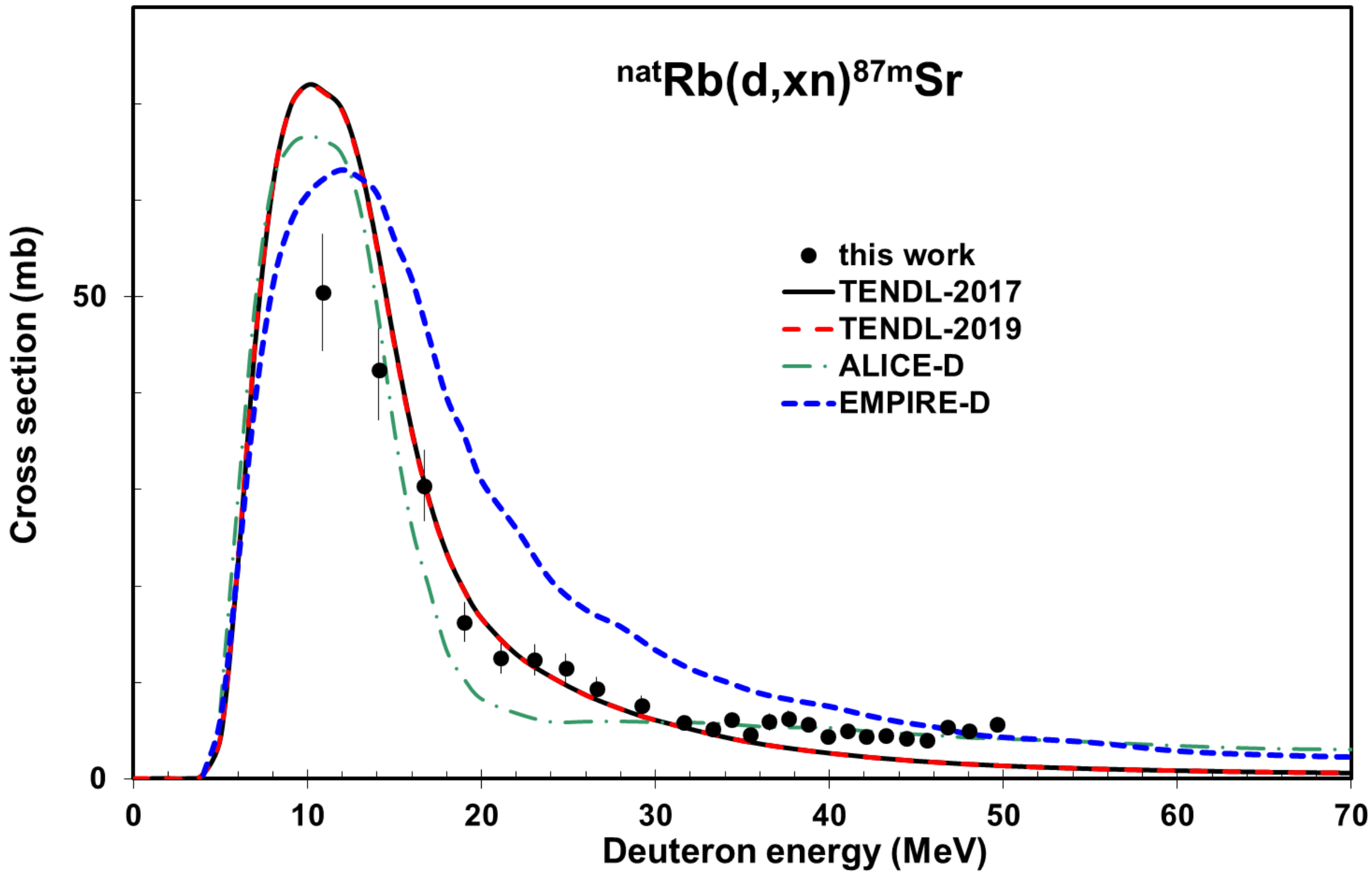

Fig. 2 Excitation function (indicative) of the $^{nat}$Rb(d,xn)$^{87m}$Sr reaction in comparison with theoretical results.

4.1.2 *$^{nat}$Rb(d,xn)$^{85m}$Sr*

The radionuclide $^{85}$Sr has a longer-lived ground state ($T_{1/2}$ = 64.849 d) and a shorter-lived isomeric state ($T_{1/2}$ = 67.63 min) decaying partly to the ground state. The excitation function for production of the metastable state is shown in Fig. 3. There is a good agreement with the TENDL-2019 (TENDL-2017) predictions at lower energies, representing the $^{85}$Rb(d,2n) contribution. But around the weak maximum of the $^{87}$Rb(d,4n) reaction the predictions of the three codes are significantly different. The both versions of the TENDL gave the same results.

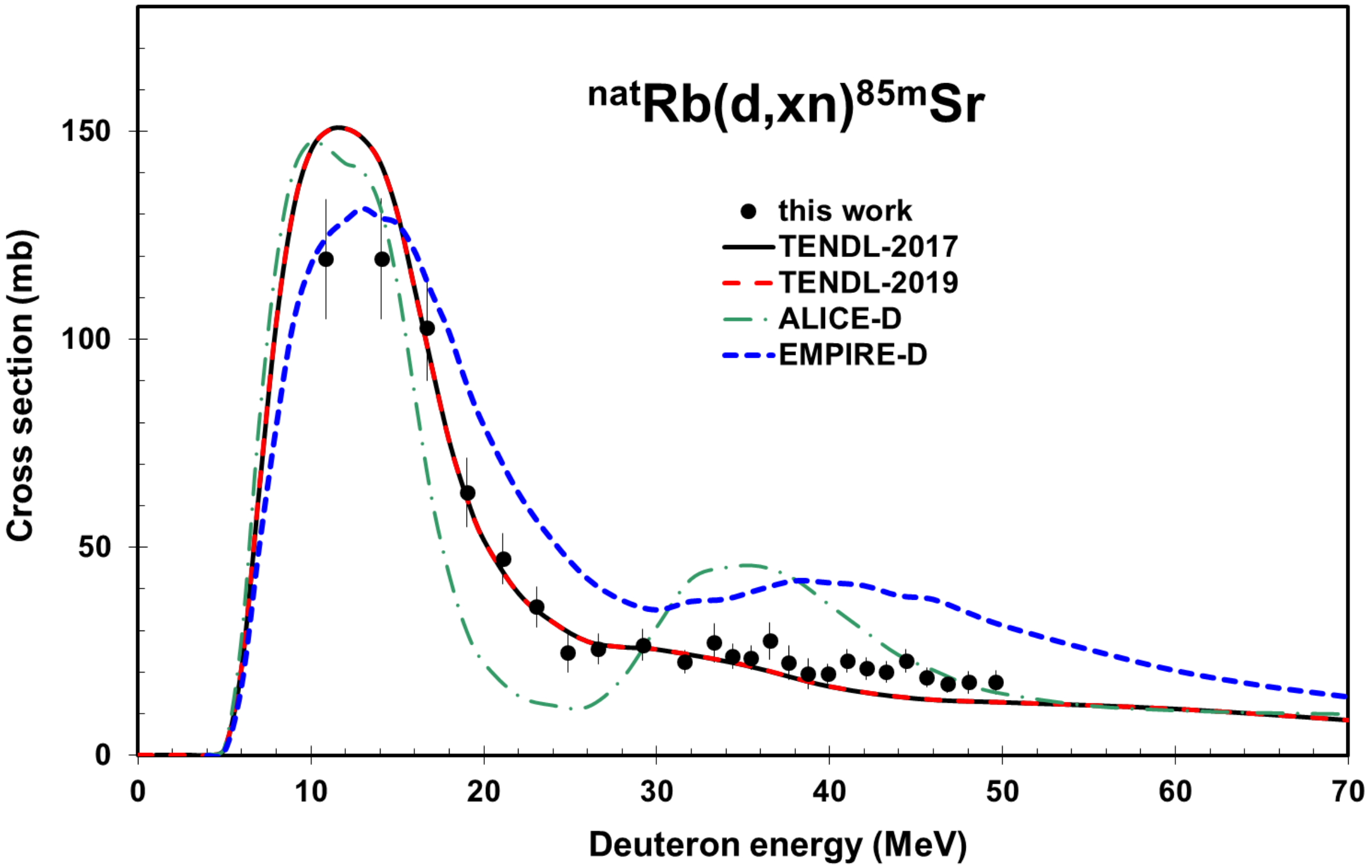

Fig. 3 Excitation function of the $^{nat}$Rb(d,xn)$^{85m}$Sr reaction in comparison with theoretical results.

4.1.3 *$^{nat}$Rb(d,xn)$^{85g}$Sr (m+)*

To determine the activity of $^{85g}$Sr ($T_{1/2}$ = 64.849 d) with gamma-spectrometry, it is necessary to separate the 514 keV gamma-line from the strong annihilation peak. The measured cross sections include the complete decay contribution from the short-lived metastable state ($T_{1/2}$ = 67.63 min, IT 83.6 %) (see above). Comparing with code results the magnitudes are similar, but the shapes differ significantly (Fig. 4).

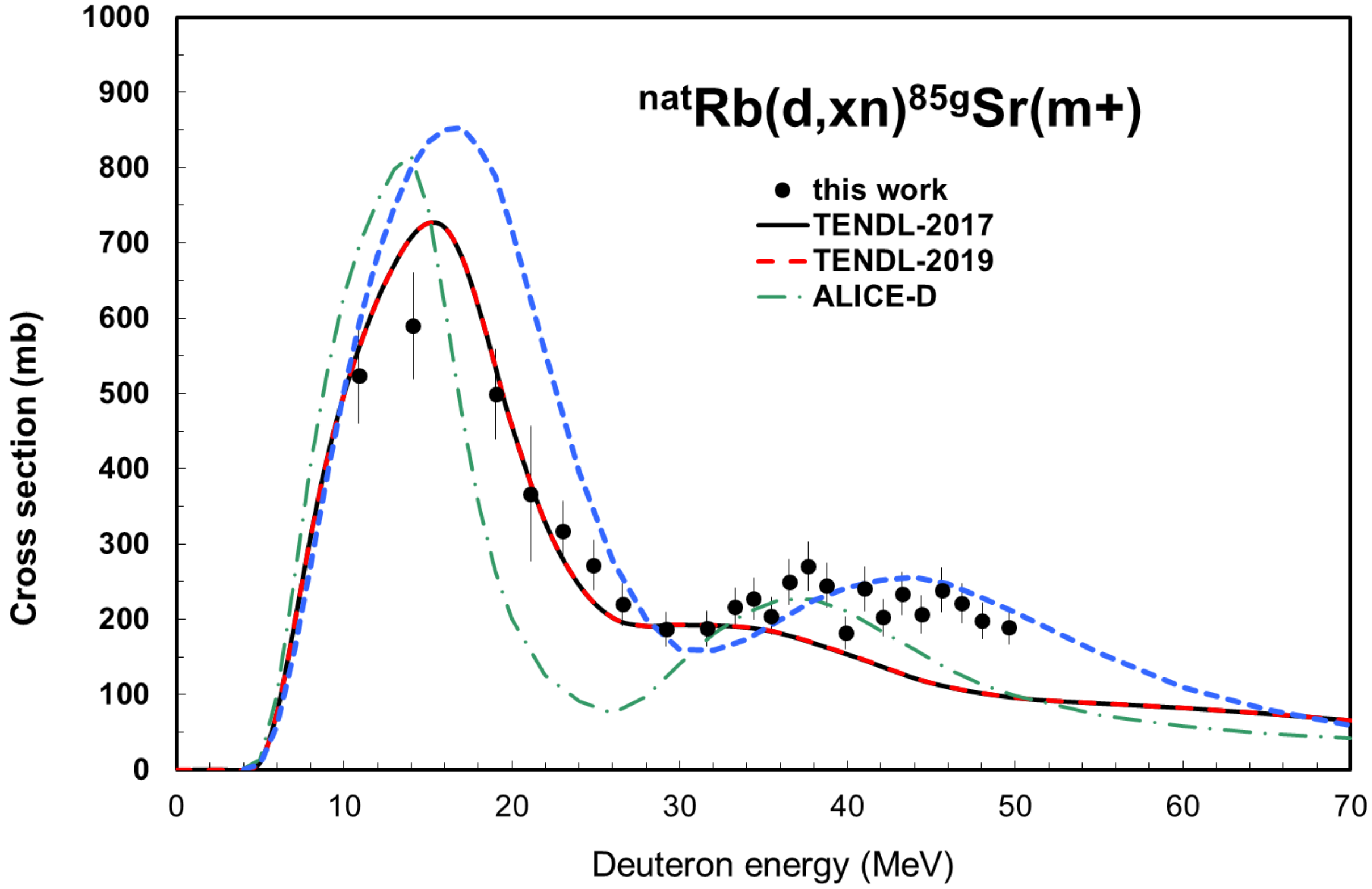

Fig. 4 Excitation function of the $^{nat}$Rb(d,xn)$^{85g}$Sr(m+) reaction in comparison with theoretical results.

### 4.1.4 *$^{nat}$Rb(d,xn)$^{83g}$Sr(m+)*

As it is obvious from Table 2, the very short-lived isomeric state of $^{83}$Sr ($T_{1/2}$ = 4.95 s, IT: 100 %) decayed completely to the ground state ($T_{1/2}$ = 32.41 h) before the spectra measurements started. The experimental and theoretical data for the excitation curve of $^{83g}$Sr(m+) are systematically shifted in energy and differ in maximum cross section value (Fig. 5).

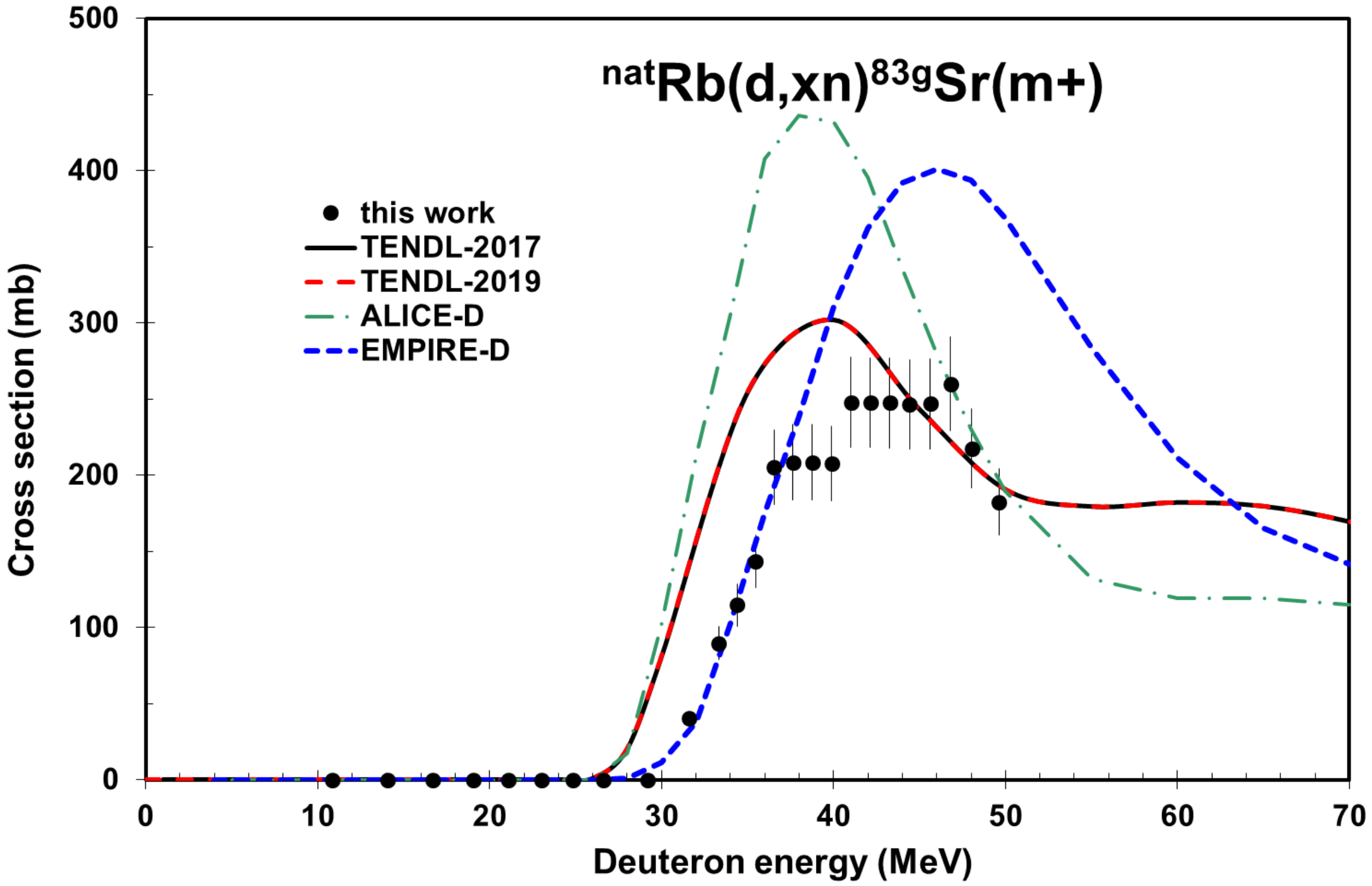

Fig. 5 Excitation function of the $^{nat}Rb(d,xn)^{83g}Sr(m+)$ reaction in comparison with theoretical results.

### 4.1.5 $^{nat}$*Rb(d,xn)*$^{82}$*Sr*

The longer-lived radionuclide $^{82}Sr$ ($T_{1/2}$ = 25.35 d) has no gamma-lines. The activity was measured through the 776 keV gamma-line of its daughter $^{82g}Rb$ ($T_{1/2}$ = 1.2375 min). The comparison of the experimental and the theoretical data (in our experimental conditions only contribution of $^{85}Rb(d,5n)^{82}Sr$) are presented in Fig. 6, showing an excellent agreement with EMPIRE-D while an energy shift with the other codes appear.

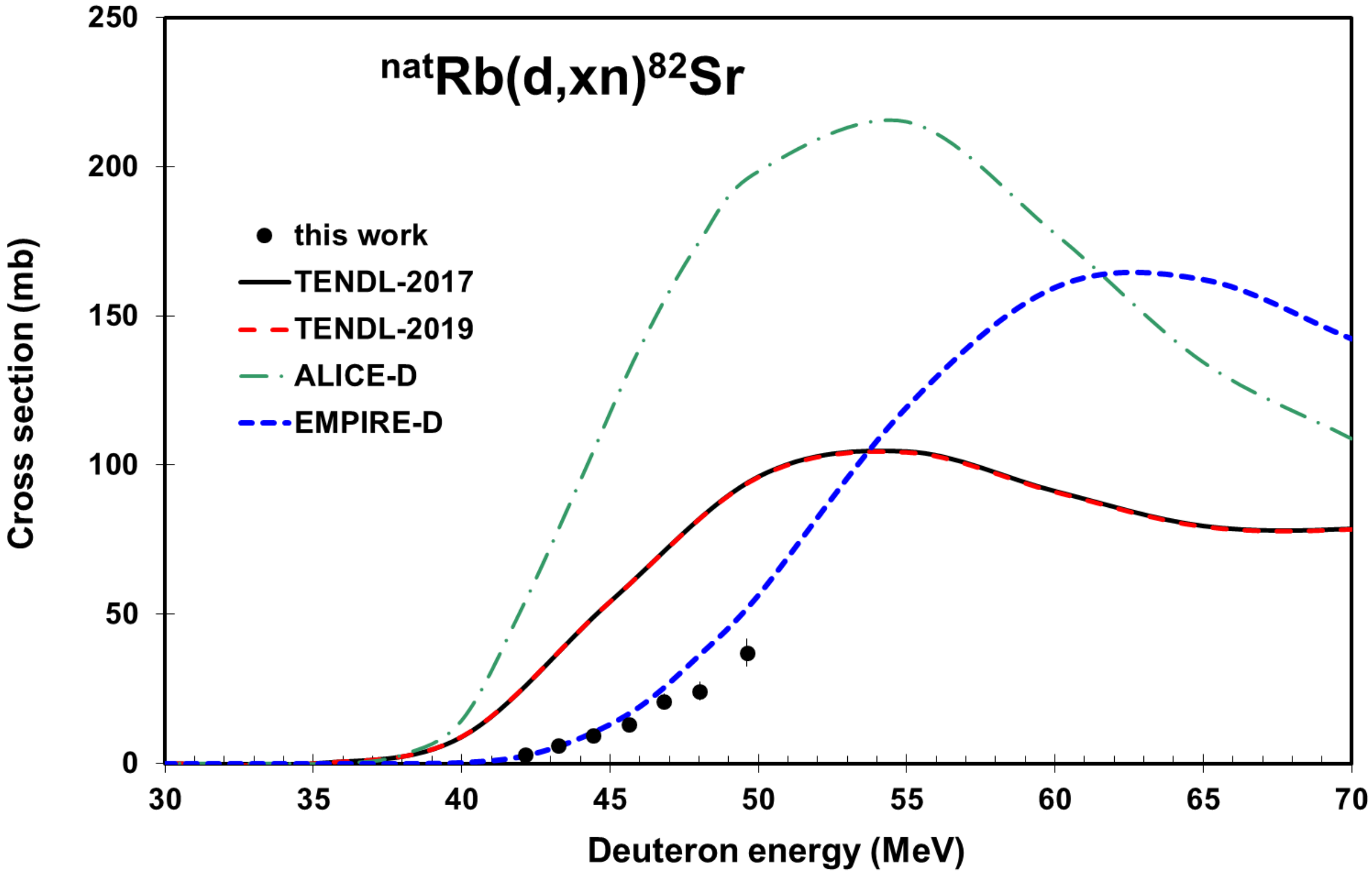

Fig. 6 Excitation function of the $^{nat}Rb(d,xn)^{82}Sr$ reaction in comparison with theoretical results.

4.1.6 Calculations for strontium isotopes

The statistical codes EMPIRE-D and TALYS describe rather well the near-threshold parts of the neutron emission reactions. The ALICE-D results are less satisfactory for this region through overly simplified modeling of the low-lying discrete levels of the considered nuclei. For calculations of the cross sections in the region of maximums, a ratio of the level densities of competing reaction channels is crucial. The existing differences in calculations reflect differences of the default level-density parameters of the corresponding codes. Discrepancies between calculations with experiment can be eliminated by the appropriate adjustments of parameters. For the $^{nat}Rb(d,xn)^{83g}Sr(m+)$ and $^{nat}Rb(d,xn)^{82}Sr$ reactions the EMPIRE-D results look preferable over other ones.

## 4.2 Excitation functions for production of rubidium radioisotopes

The radioisotopes of rubidium are produced via direct (d,pxn) reactions and through the ε and/or $\beta^+$ decay of Sr parent radioisotopes.

4.2.1 $^{nat}Rb(d,x)^{86g}Rb$ *(m+)*

The longer-lived ground state $^{86}$Rb ($T_{1/2}$ = 18.642 d) is produced via direct (d,pxn) reaction on $^{85,87}$Rb. The measured data are so called (m+), including the complete decay of the isomeric state ($T_{1/2}$ = 1.017 min, IT 100 %) (Fig. 7). The description of $^{87}$Rb(d,p) part is not satisfactory in all model codes. The experimental data are in good agreement with the systematics for a (d,p) reaction in this mass region [20].

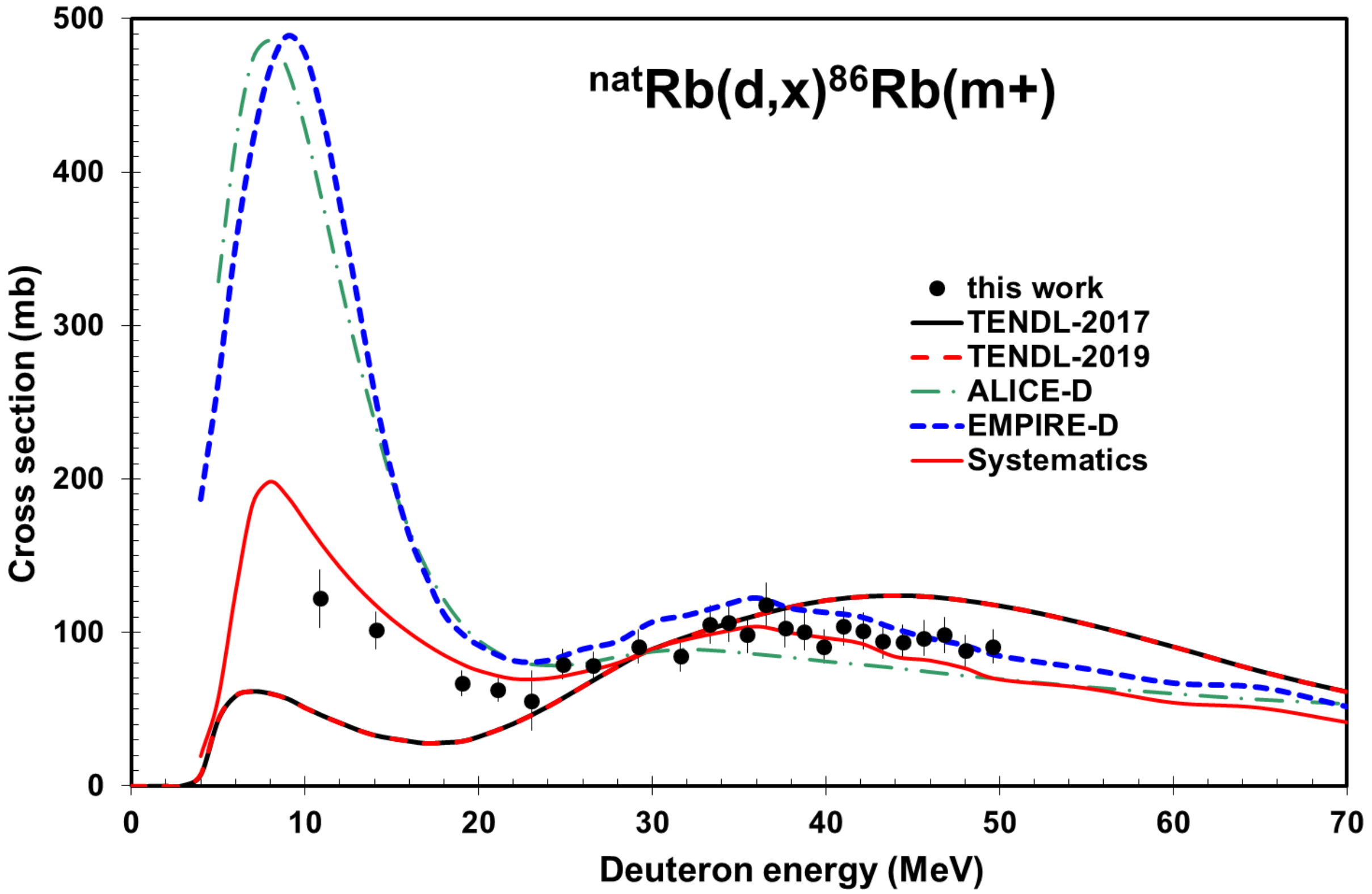

Fig. 7 Excitation function of the $^{nat}$Rb(d,x)$^{86}$Rb(m+) reaction in comparison with theoretical results.

**4.2.2** $^{nat}Rb(d,x)^{84g}Rb(m+)$

The $^{84g}$Rb ($T_{1/2}$ = 32.82 d) and $^{84m}$Rb ($T_{1/2}$ = 20.26 min, IT 100 %) isomers are produced only directly on $^{nat}$Rb targets via $^{85}$Rb(d,p2n) and $^{87}$Rb(d,p4n) reactions. The cross sections for production of $^{84g}$Rb were deduced from spectra measured after complete isomeric decay of the short-lived metastable state. Among the model predictions only the EMPIRE-D descriptions are acceptable over the energy region studied (Fig. 8).

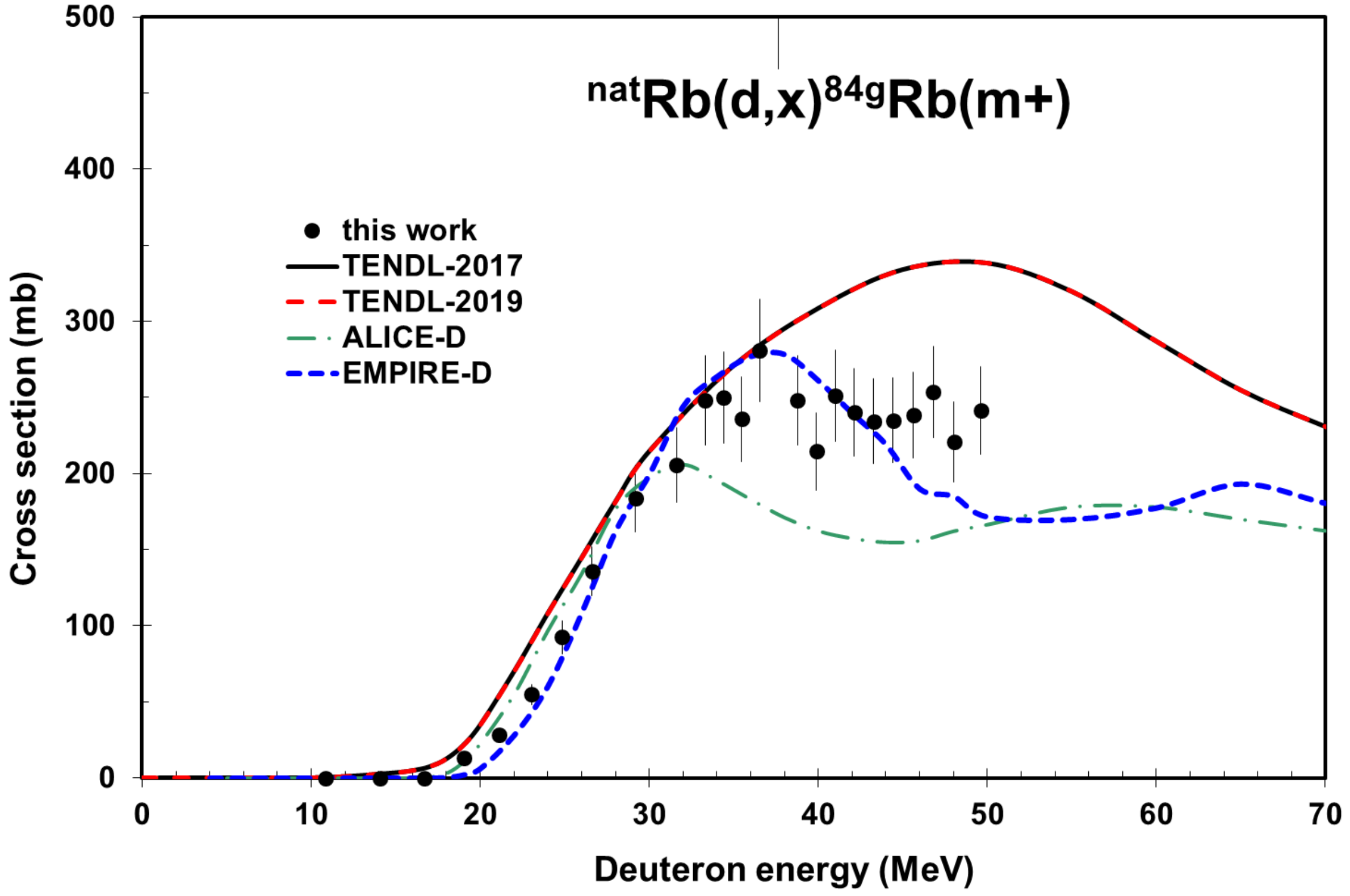

Fig. 8 Excitation function (indicative) of the $^{nat}Rb(d,x)^{84g}Rb(m+)$ reaction in comparison with theoretical results.

### 4.2.3 $^{nat}$*Rb(d,x)*$^{83}$*Rb(cum)*

The measured production cross section of $^{83}Rb$ ($T_{1/2}$ = 86.2 d) are cumulative. It contains the complete decay of its $^{83}Sr$ ($T_{1/2}$ = 32.41 h, ε:100 %) parent isotope (Fig. 9). The model predictions are very different and cannot even describe rather well the cumulative maximum. EMPIRE-D is acceptable up to 38 MeV, and the cumulative TENDL version give the better results above this energy.

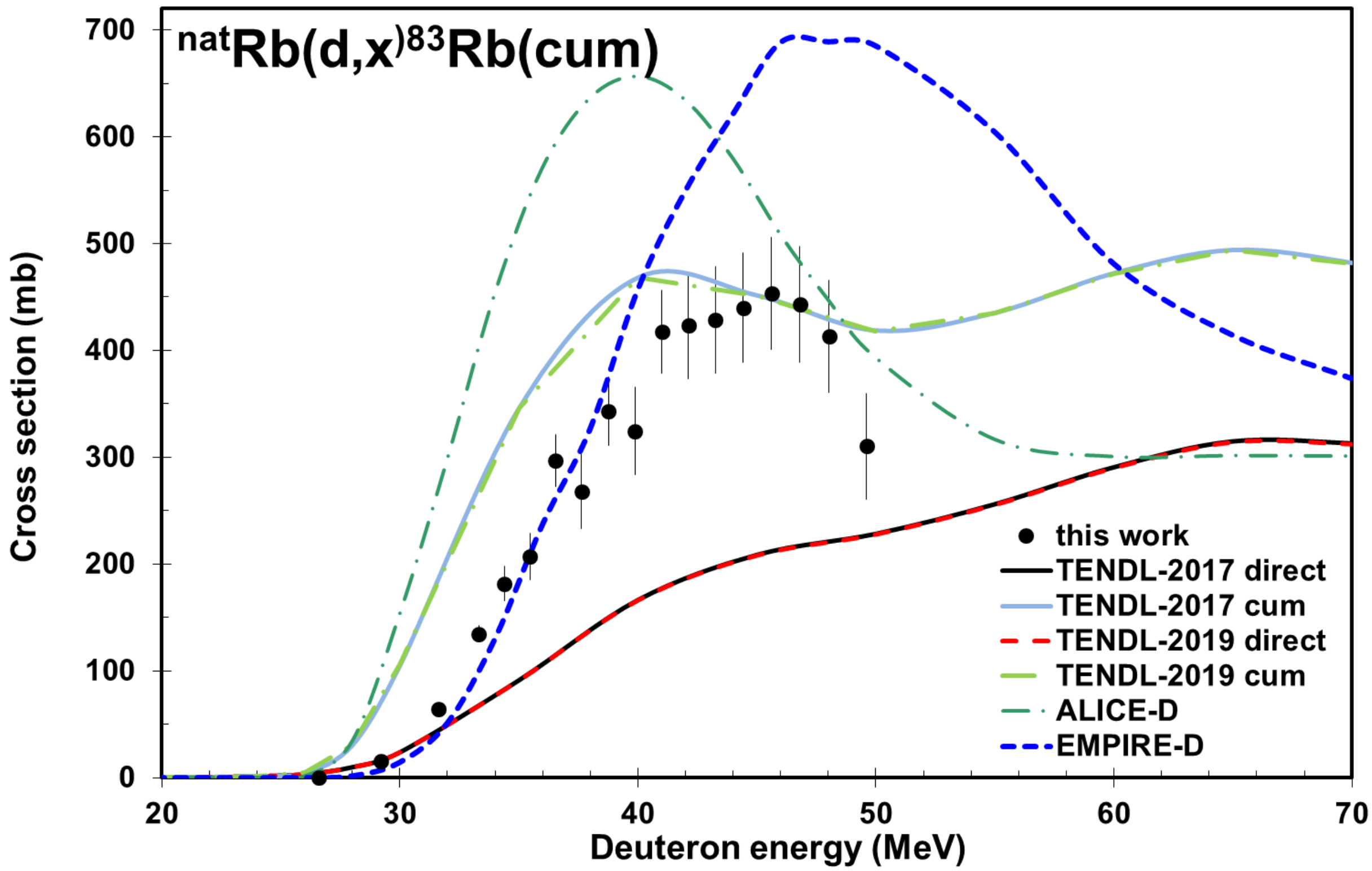

Fig. 9 Excitation function of the $^{nat}Rb(d,x)^{83}Rb(cum)$ reaction in comparison with theoretical results.

### 4.2.4 *$^{nat}Rb(d,p4n)^{82m}Rb$*

The metastable state $^{82m}Rb$ ($T_{1/2}$ = 6.472 h) is produced here directly only via the $^{85}Rb(d,p4n)$ reaction (Q = -32.41 MeV). The experimental data, with high practical threshold, are described more satisfactory by the ALICE-D prediction (Fig. 10).

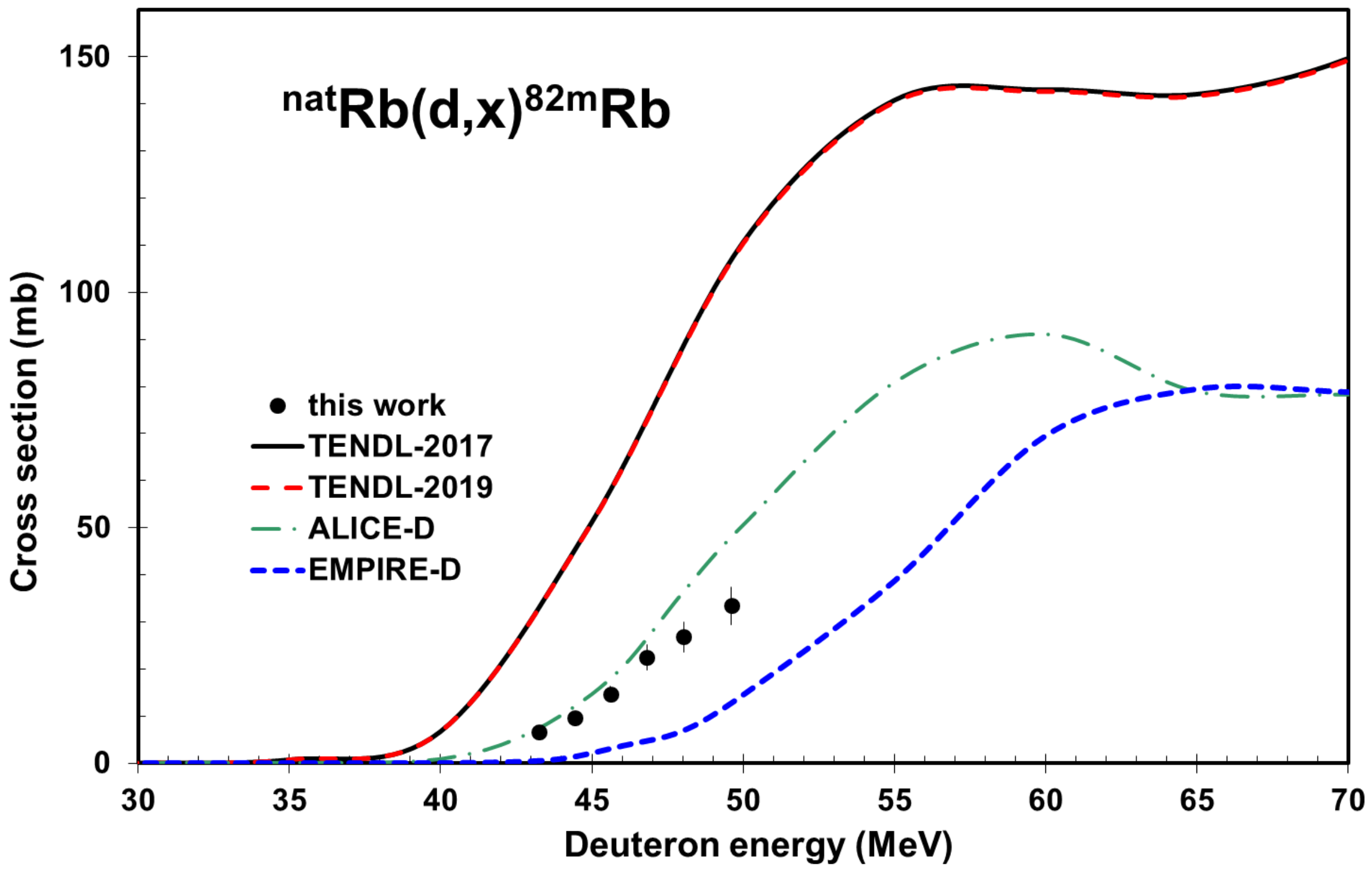

Fig. 10 Excitation function of the $^{nat}Rb(d,x)^{82m}Rb$ reaction in comparison with theoretical results.

4.2.5 Calculations for rubidium isotopes

The general comments on the (d, pxn) reactions are the same as those discussed in section 4.1.6 above. However, a special consideration is required for the (d, p) reaction. It is well known that this reaction is dominated by the mechanism of a direct breakup without the formation of any intermediate pre-equilibrium state of the nucleus. A satisfactory description of the experimental data on this reaction can be achieved only on the basis of the phenomenological systematics proposed in [21]. For the $^{nat}Rb(d,x)^{84g}Rb(m+)$ reaction the TENDL-2019 results look a little preferable over the EMPIRE-D ones and for the $^{nat}Rb(d,xn)^{83}Rb(cum)$ reaction we observed an inverse case. For the $^{nat}Rb(d,xn)^{82m}Rb$ reaction the ALICE-D results are certainly preferable, but that seems rather accidental result. Description of experimental data can be surely improved for each reaction by the appropriate adjustments of the nuclear level density parameters.

## 4.3 Excitation functions for production of krypton radioisotopes

4.3.*1* $^{nat}$*Rb(d,x)*$^{85m}$*Kr*

The metastable state $^{85m}$Kr ($T_{1/2}$ = 4.480 h) is produced directly by $^{85}$Rb(d,2p) and $^{87}$Rb(d,α) reactions, both with low thresholds. The $^{85m}$Kr has common gamma-lines with $^{85}$Sr (both decay to stable $^{85}$Rb). The presented data were corrected for contribution of the $^{85}$Sr in the measured gamma-spectra. The 151 keV gamma-line was present only in the first(early) series of gamma spectra measurements and with low statistics. We had possibility to deduce the cross section data for $^{85m}$Kr via subtracting the contribution of the $^{85m}$Sr. The contribution was significant and taking into account that the cross section of the $^{85m}$Kr is low, the statistics of the separated 151 of the $^{85m}$Kr was poor. The experimental data, slowly rising as a result of the combination of the two contributing reactions without pronounced maximum in the studied energy region, and their curve is described acceptable well by the used model codes (Fig. 11).

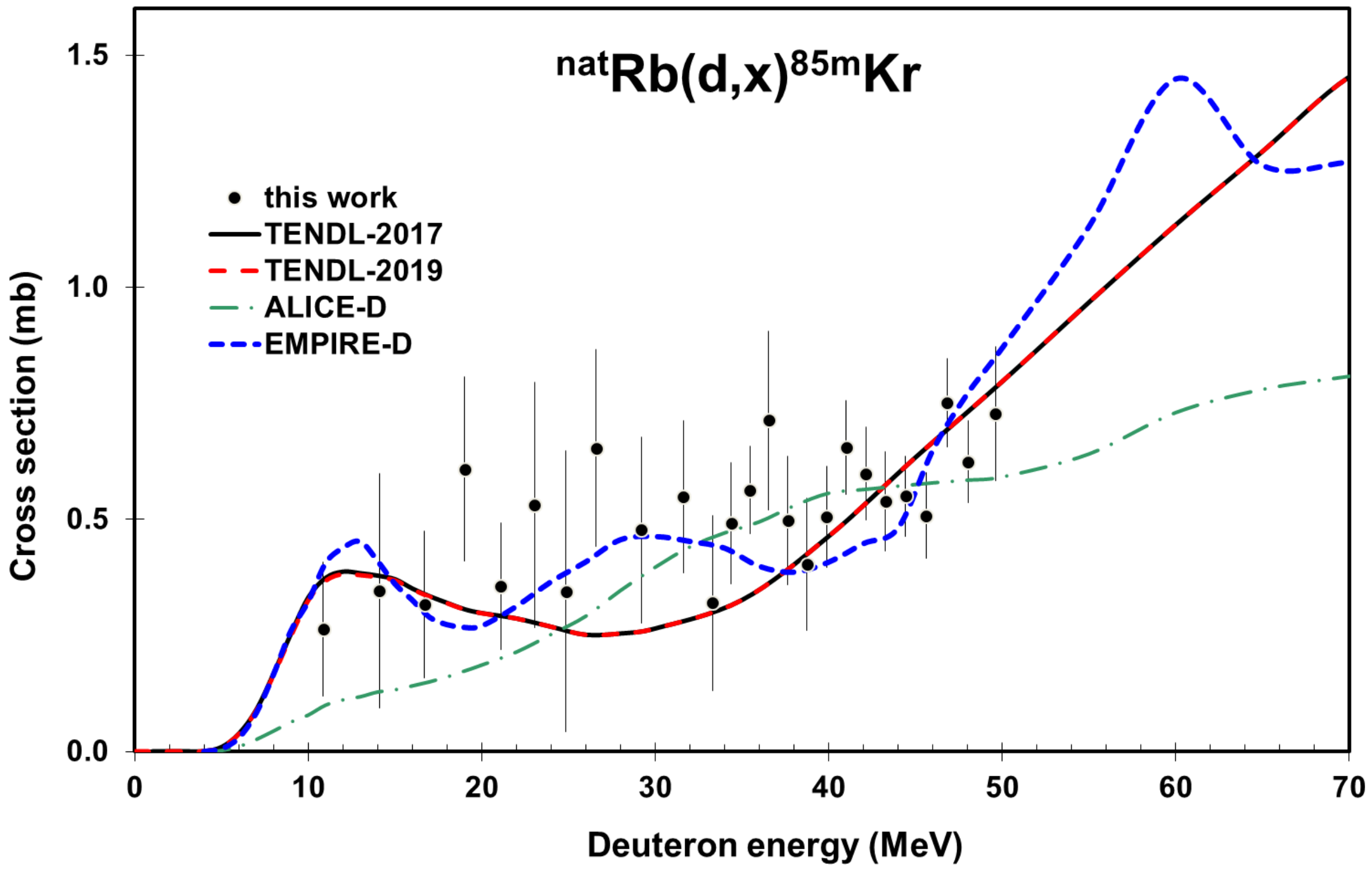

Fig.11 Excitation function of the $^{nat}$Rb(d,x)$^{85m}$Kr reaction in comparison with theoretical results.

**4.4 The thick target and integral yields of the investigated radioisotopes**

The thick target yields as a function of the bombarding energy, calculated from integration of a fit to our experimental excitation functions (Eq. 3), are shown in Figs. 12 and 13 compared to values obtained in direct thick target measurements [1]. All thick target and integral yields are deduced for metallic Rb targets of natural composition. The integral yields represent so called physical yields (see [11] and [12]). The only experimental values are for $^{85g}$Sr and $^{86g,84g}$Rb available [1] and shows acceptable agreement with our curve at 22 MeV deuteron energy.

$$Y(E) = \int_{E_{th}}^{E} F\sigma(E)dE \qquad \text{Equation 3}$$

Where $E_{th}$ is the threshold energy of the nuclear reaction in question, and $F$ is the conversion constant, which contains all the constants as density, molar weight, decay constant, electron charge, Avogadro's number, etc. and the conversion factors necessary to get the integral yield (*Y(E)*) in proper units. Instead of integral along a numerical integration was performed along depth with $\Delta x(E,dE)$, where $dE$ is equidistant, and in such a way $\Delta x$ is not.

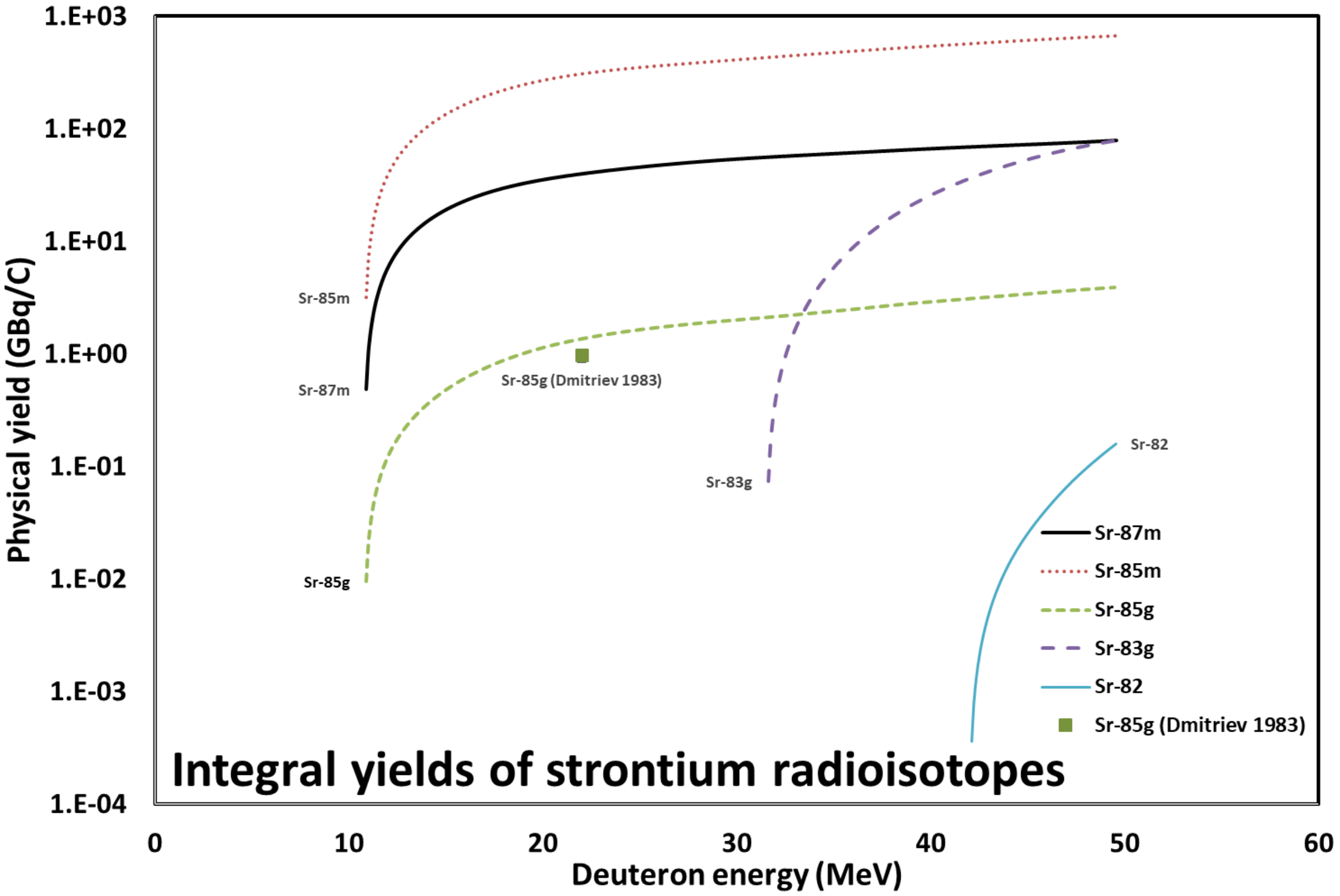

Fig. 12 $^{nat}$Rb(d,x) integral yields for production of strontium radioisotopes.

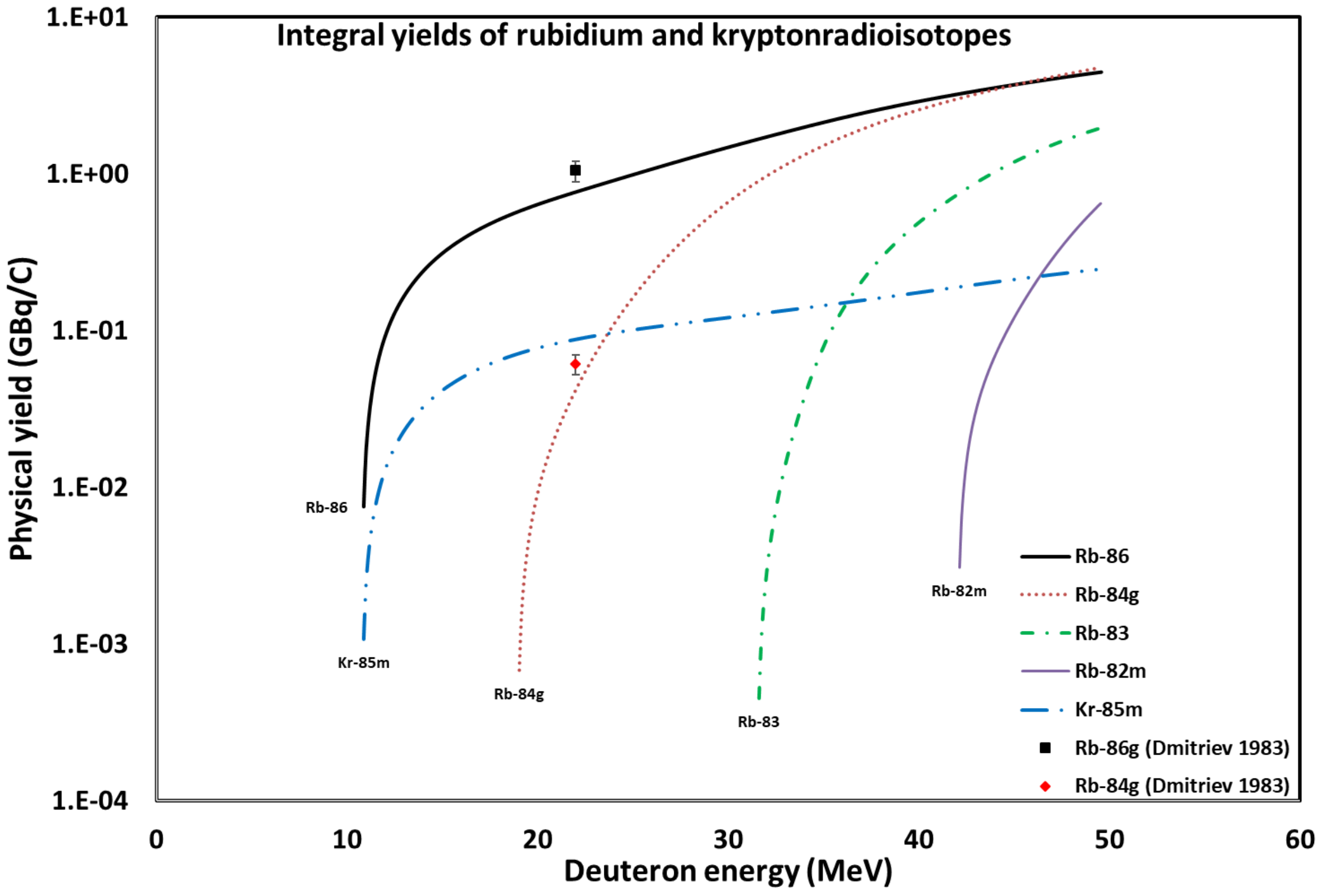

Fig. 13 $^{nat}$Rb(d,x) integral yields for production of rubidium and krypton radioisotopes.

## 5. Applications of deuteron induced reactions on rubidium for nuclear medicine

Many investigated activation products are relevant for nuclear medicine ($^{87m}$Sr, $^{85g}$Sr, $^{83g}$Sr, $^{82}$Sr ($^{82}$Rb), $^{82m}$Rb, $^{85m}$Kr). Through the presently investigated reactions, the production of $^{82m}$Rb is carrier added and the production yields of $^{85m}$Kr and $^{82}$Sr (in the investigated energy range) are very low. Only production of the $^{87m}$Sr, $^{85}$Sr, $^{83}$Sr and $^{82}$Sr radioisotopes are hence discussed shortly in more detail. More detailed discussion, including production parameters (yield, impurities, targetry, etc.) will be prepared and published in a dedicated journal.

### 5.1 Production of $^{87m}$Sr

$^{87m}$Sr (2.815 h) is used in skeletal SPECT (Single Photon Emission Computer Tomography) imaging for diagnosis of diseases. This radionuclide can be produced directly: by proton and deuteron induced reaction on rubidium: $^{nat}$Rb(p,x)$^{87m}$Sr [22], $^{87}$Rb(p,x)$^{87m}$Sr [23-25],

$^{nat}$Rb(d,x)$^{87m}$Sr (this work) or by alpha and $^{3}$He induced reactions on krypton: $^{nat}$Kr(α,x)$^{87m}$Sr [26], $^{nat}$Kr($^{3}$He,n)$^{87m}$Sr [26], or indirectly: through $^{87}$Y (13.37 h)/$^{87m}$Sr generator by alpha and $^{3}$He induced reactions on rubidium: $^{nat}$Rb(α,x)$^{87m}$Y [27, 28], $^{85}$Rb(α,x)$^{87m}$Y [29, 30], $^{85}$Rb(α,x)$^{87g}$Y [29, 30], $^{85}$Rb(α,x)$^{87}$Y [28, 30-33], $^{nat}$Rb($^{3}$He,x)$^{87m}$Y [2, 27], $^{nat}$Rb($^{3}$He,x)$^{87g}$Y [2] or proton induced reaction on enriched $^{87}$Sr target [29, 34-37], proton induced reaction on $^{88}$Sr target [29, 38-40] or proton induced reaction on natural strontium target [41-43] or deuteron induced reactions on $^{nat}$Sr [44, 45]. In practical medical isotope production, the role of the $^{3}$He induced reaction is minimal due to lack of accelerator and high price of $^{3}$He and the high stopping power. The situation is practically same in case of alpha particle induced reactions, except the price of $^{4}$He gas. Alpha particles are used only when no low energy possibility with protons and deuterons is available ($^{211}$At). In practice proton induced reactions are the favorites due to the large cross sections, availability of high intensity beams, and lower stopping power. But during our systematic study of deuteron induced reactions it was shown that some isotopes cannot be produced by protons, but they are available with deuterons via (d,n) and (d,p) reactions, and at medium and heavy mass targets the production yields of (d,2n) are significantly higher comparing to (p,n).

### 5.2 Production of $^{85}$Sr

The $^{85g}$Sr ($T_{1/2}$ = 64.849 d) had importance in nuclear medicine as diagnostic radioisotope, later it was proposed to use it for the endotherapy and as important disturbing impurity in the application of the $^{82}$Sr and $^{83g}$Sr. It can be produced directly or indirectly through $^{85}$Y parent. In the literature the following direct and indirect production routes have been found. Direct production routes: $^{85}$Rb(p,n)$^{85}$Sr [29, 46-48], $^{nat}$Rb(p,xn) $^{85}$Sr [22, 24, 49-53], $^{nat}$Rb(d,xn)$^{85}$Sr (this work), $^{nat}$Kr(α,n)$^{85m}$Sr [26, 54], $^{nat}$Kr(α,n)$^{85g}$Sr [26, 54], $^{nat}$Kr($^{3}$He,n)$^{85m}$Sr [26, 54], $^{nat}$Kr($^{3}$He,n)$^{85g}$Sr [26, 54]. Indirect production routes: $^{86}$Sr(p,2n)$^{85}$Y [2, 29], $^{nat}$Sr(d,xn)$^{85}$Y [45].

### 5.3 Production of $^{83}$Sr

The $^{83g}$Sr is a positron emitter analog of the $\beta^-$ emitting $^{89}$Sr ($T_{1/2}$ = 50.5 d) used for endotherapy.

The following production routes have been investigated in the literature: $^{nat}Rb(p,xn)^{83}Sr$ [22, 24, 49], $^{85}Rb(p,3n)^{83}Sr$ [29, 48, 55], $^{nat}Rb(d,xn)^{83}Sr$ [this work], $^{82}Kr(^{3}He,n)^{83}Sr$ [26, 54], $^{nat}Kr(\alpha,xn)^{83}Sr$ [26, 54], $^{nat}Kr(^{3}He,3n)^{83}Sr$ [26, 54].

### 5.4 $^{82}$Sr production

$^{82}Sr(^{82}Rb)$ is widely used in myocardial perfusion imaging in Positron Emission Tomography (PET). The following production routes have been investigated: $^{85}Rb(p,4n)^{82}Sr$ [22, 48, 49, 53, 56], $^{nat}Rb(p,xn)^{82}Sr$ [22, 48, 49, 52, 53, 55-57], $^{nat}Rb(d,xn)^{82}Sr$ [this work], $^{82}Kr(\alpha,4n)^{82}Sr$ [26], $^{83}Kr(\alpha,5n)^{82}Sr$ [58], $^{82}Kr(^{3}He,3n)^{82}Sr$ [26], $^{83}Kr(^{3}He,4n)^{82}Sr$ [26], $^{nat}Kr(^{3}He,xn)^{82}Sr$ [26, 54], $^{nat}Kr(\alpha,xn)^{82}Sr$ [26, 54].

## 6. Summary and conclusions

Activation cross sections of the $^{nat}Rb(d,xn)^{87m,85m,85g,83,82}Sr$, $^{nat}Rb(d,x)^{86,84,83,82m}Rb$ and $^{nat}Rb(d,x)^{85m}Kr$ nuclear reactions were measured for the first time up to 50 MeV relative to well documented $^{27}Al(d,x)^{22,24}Na$ monitor reactions. Model calculations were done with the ALICE-IPPE-D and EMPIRE D code and completed with the TALYS results from TENDL-2019 library. The model calculations predict with varying success the shape and the absolute values of the experimental data. Physical integral yields were deduced for every studied radioisotope and compared with the literature data for production of $^{84,86}$Rb and $^{85}$Sr. The status of cross section and yield data for production of some medically important radioisotopes is also discussed.

The obtained experimental data also provide a basis for improved model calculations and for applications in radioisotope production.

## Acknowledgements

This work was done in the frame of MTA-FWO (Vlaanderen) research projects. The authors acknowledge the support of research projects and of their respective institutions in providing the materials and the facilities for this work. This work was also partly supported (F. Ditroi) by IAEA RER Project 1020 and by IAEA CRP F22069.

Table 3. Cross sections of deuteron induced reactions on rubidium for production of strontium radioisotopes

| | | $^{87m}Sr$ | | $^{85m}Sr$ | | $^{85g}Sr$ | | $^{83}Sr$ | | $^{82}Sr$ | |
|---|---|---|---|---|---|---|---|---|---|---|---|
| **E** | **dE** | **σ** | **dσ** | **σ** | **dσ** | **σ** | **dσ** | **σ** | **dσ** | **σ** | **dσ** |
| **MeV** | | **mb** | | | | | | | | | |
| 49.62 | 0.30 | 5.60 | 0.74 | 17.56 | 2.81 | 189.29 | 22.61 | 182.26 | 21.78 | 37.11 | 8.3 |
| 48.03 | 0.32 | 5.01 | 0.62 | 17.57 | 2.71 | 198.02 | 23.93 | 217.62 | 25.96 | 24.25 | 5.4 |
| 46.81 | 0.34 | 5.36 | 0.67 | 17.18 | 2.11 | 221.82 | 26.65 | 260.03 | 31.05 | 20.74 | 4.6 |
| 45.62 | 0.36 | 3.98 | 0.50 | 18.78 | 2.27 | 239.35 | 29.04 | 246.76 | 29.45 | 13.11 | 2.9 |
| 44.42 | 0.38 | 4.17 | 0.51 | 22.72 | 2.75 | 206.54 | 25.11 | 246.54 | 29.42 | 9.40 | 2.1 |
| 43.27 | 0.40 | 4.48 | 0.57 | 20.15 | 2.53 | 234.17 | 28.39 | 247.41 | 29.53 | 6.01 | 1.3 |
| 42.13 | 0.43 | 4.36 | 0.58 | 20.84 | 2.66 | 202.69 | 24.16 | 247.71 | 29.57 | 3.00 | 0.7 |
| 41.02 | 0.45 | 4.96 | 0.62 | 22.74 | 2.80 | 240.69 | 29.18 | 247.79 | 29.58 | | |
| 39.88 | 0.48 | 4.43 | 0.56 | 19.57 | 2.37 | 182.17 | 21.96 | 207.57 | 24.80 | | |
| 38.76 | 0.51 | 5.66 | 0.73 | 19.57 | 3.66 | 245.48 | 30.03 | 208.31 | 24.87 | | |
| 37.64 | 0.54 | 6.24 | 0.85 | 22.28 | 4.17 | 270.34 | 32.78 | 208.51 | 24.92 | | |
| 36.54 | 0.57 | 5.91 | 0.85 | 27.52 | 4.38 | 249.89 | 29.82 | 205.16 | 24.56 | | |
| 35.46 | 0.61 | 4.61 | 0.57 | 23.44 | 2.83 | 204.50 | 24.69 | 143.45 | 17.20 | | |
| 34.40 | 0.65 | 6.10 | 0.77 | 23.80 | 3.02 | 227.85 | 27.49 | 114.70 | 13.78 | | |
| 33.31 | 0.69 | 5.17 | 0.76 | 27.04 | 4.62 | 216.11 | 26.14 | 89.73 | 10.84 | | |
| 31.61 | 0.73 | 5.81 | 0.74 | 22.43 | 2.80 | 187.90 | 23.05 | 40.06 | 4.97 | | |
| 29.20 | 0.77 | 7.60 | 0.97 | 26.52 | 3.76 | 187.31 | 22.78 | | | | |
| 26.61 | 0.82 | 9.34 | 1.18 | 25.55 | 3.67 | 219.83 | 28.69 | | | | |
| 24.85 | 0.87 | 11.46 | 1.52 | 24.69 | 4.80 | 272.35 | 33.15 | | | | |
| 23.04 | 0.92 | 12.34 | 1.60 | 35.66 | 4.92 | 317.21 | 40.44 | | | | |
| 21.11 | 0.98 | 12.49 | 1.55 | 47.21 | 6.13 | 366.95 | 90.05 | | | | |
| 19.03 | 1.04 | 16.26 | 2.05 | 63.20 | 8.34 | 499.45 | 59.72 | | | | |
| 16.72 | 1.10 | 30.39 | 3.69 | 102.82 | 12.75 | | | | | | |
| 14.09 | 1.17 | 42.36 | 5.10 | 119.40 | 14.52 | 589.95 | 70.65 | | | | |
| 10.87 | 1.24 | 50.48 | 6.06 | 119.29 | 14.32 | 523.97 | 63.35 | | | | |

Table 4. Cross sections of deuteron induced reactions on rubidium for production of rubidium and krypton radioisotopes

| | | $^{86}$Rb | | $^{84}$Rb | | $^{83}$Rb | | $^{82m}$Rb | | $^{85m}$Kr | |
|---|---|---|---|---|---|---|---|---|---|---|---|
| E | dE | σ | dσ | σ | dσ | σ | dσ | σ | dσ | σ | dσ |
| MeV | | mb | | | | | | | | | |
| 49.62 | 0.30 | 90.81 | 10.93 | 241.54 | 28.79 | 413.06 | 49.25 | 33.47 | 4.02 | 0.73 | 0.14 |
| 48.03 | 0.32 | 87.96 | 10.58 | 220.64 | 26.30 | 443.13 | 52.83 | 26.83 | 3.21 | 0.62 | 0.09 |
| 46.81 | 0.34 | 98.31 | 11.79 | 253.54 | 30.22 | 453.14 | 54.02 | 22.46 | 2.70 | 0.75 | 0.10 |
| 45.62 | 0.36 | 96.42 | 11.55 | 238.48 | 28.42 | 439.96 | 52.44 | 14.63 | 1.76 | 0.51 | 0.09 |
| 44.42 | 0.38 | 93.85 | 11.21 | 234.82 | 27.99 | 428.60 | 51.12 | 9.60 | 1.17 | 0.55 | 0.09 |
| 43.27 | 0.40 | 94.55 | 11.36 | 234.30 | 27.93 | 423.26 | 50.46 | 6.55 | 0.81 | 0.54 | 0.11 |
| 42.13 | 0.43 | 100.98 | 12.09 | 240.26 | 28.64 | 417.62 | 49.82 | 7.51 | 0.90 | 0.60 | 0.10 |
| 41.02 | 0.45 | 104.06 | 12.54 | 251.25 | 29.96 | 324.41 | 38.71 | | | 0.65 | 0.10 |
| 39.88 | 0.48 | 90.83 | 10.94 | 214.57 | 25.58 | 343.34 | 41.00 | | | 0.51 | 0.11 |
| 38.76 | 0.51 | 100.56 | 12.16 | 247.92 | 29.57 | 268.18 | 32.01 | | | 0.40 | 0.14 |
| 37.64 | 0.54 | 102.87 | 12.57 | 528.61 | 63.01 | 296.76 | 35.39 | | | 0.50 | 0.14 |
| 36.54 | 0.57 | 118.09 | 14.15 | 280.81 | 33.47 | 206.82 | 24.67 | | | 0.71 | 0.19 |
| 35.46 | 0.61 | 98.32 | 11.79 | 235.69 | 28.09 | 181.66 | 21.68 | | | 0.56 | 0.09 |
| 34.40 | 0.65 | 106.53 | 12.79 | 249.95 | 29.80 | 134.54 | 16.07 | | | 0.49 | 0.13 |
| 33.31 | 0.69 | 105.43 | 12.65 | 248.19 | 29.58 | 64.38 | 7.71 | | | 0.32 | 0.19 |
| 31.61 | 0.73 | 84.56 | 10.15 | 205.45 | 24.49 | 15.75 | 2.00 | | | 0.55 | 0.16 |
| 29.20 | 0.77 | 90.80 | 10.91 | 183.66 | 21.90 | | | | | 0.48 | 0.20 |
| 26.61 | 0.82 | 78.20 | 9.32 | 135.59 | 16.18 | | | | | 0.65 | 0.21 |
| 24.85 | 0.87 | 79.22 | 9.68 | 92.41 | 11.04 | | | | | 0.35 | 0.30 |
| 23.04 | 0.92 | 55.63 | 19.69 | 54.90 | 6.58 | | | | | 0.53 | 0.26 |
| 21.11 | 0.98 | 62.77 | 7.63 | 28.12 | 3.37 | | | | | 0.36 | 0.14 |
| 19.03 | 1.04 | 67.18 | 8.21 | 13.35 | 1.63 | | | | | 0.61 | 0.20 |
| 16.72 | 1.10 | | | | | | | | | 0.32 | 0.16 |
| 14.09 | 1.17 | 101.35 | 12.34 | | | | | | | 0.35 | 0.25 |
| 10.87 | 1.24 | 122.07 | 19.08 | | | | | | | 0.26 | 0.14 |